\documentclass[aps,prl,preprint,groupedaddress,longbibliography]{revtex4-1}

\begin{document}


\title{New approach for multiconfigurational Exchange-Correlation Functional}



\author{Yurii Dmitriev}
\email{qmech@spbu.ru}
\altaffiliation{Saint Petersburg State University, Saint Petersburg}

\author{Sergey Gusarov}
\email{gusarov@ualberta.ca}
\altaffiliation{National Research Council of Canada, Edmonton}



\date{\today}

\begin{abstract}
A new density functional theory approach based on a complete active space self-consistent field (CASSCF) reference function in Extended Koopmans' approximation is discussed. Recently, the number of generalizations of density functional theory based on a multiconfigurational CASSCF reference function with exact exchange (CASDFT) was introduced. It was shown by one of the authors (Dr. Gusarov) that such a theory could be formulated by introducing a special form of exchange-correlation potential. To take into account an active space and to avoid double counting of correlation energy the dependence from on-top pair density $P_2(\bf r)$ as a new variable was introduced. Unfortunately, this requires a deep review and reparametrization of existing functional expressions which lead to additional computational difficulties. The presented approach does not require introducing additional variables (like on-top pair density, $P_2(\bf r)$) and based on Extended Koopmans' theorem (EKT) approximation for multiconfigurational wave function within CASSCF method. 
\end{abstract}

\maketitle

\section{Introduction\label{Introduction}}

The recent success of computational chemistry in different areas of bio- and material sciences 
owes much to the development of density functional theory (DFT) methods \cite{Kohn1965,Hohenberg1964}
allowing to study complex molecular systems containing many hundreds of atoms. Moreover, the accuracy and applicability of DFT-based approximations are continuously improving due to the number of developments and generalizations allowing to study many different situations, like spin-polarized systems, time-dependent phenomena, charge transfer, etc. However, despite many new modifications, there are still a number of difficulties in using DFT to properly describe some complex systems. 
Typically such a situation arises as a result of a conflict between single-determinant mean-field representation with the multiconfiguration nature of the strongly correlated molecular systems. To overcome these difficulties a number of solutions combining multiconfigurational wave function methods with DFT have been proposed (e.g. \cite{ColleSalvetti1990,MiehlichStollSavin1997,GrafensteinCremer2000,PolletSavinLeiningerStoll2002,GusarovLindh2004,ManniGagliardiOlsenTruhlar2014}). The main problem of these theories is the double counting of non-dynamical correlation energy by density functional theory. One way to overcome this problem is to introduce an additional variable(s) which locally scale the DFT correlation to decrease it according to some additionally introduced requirements, like the ratio of active and total densities \cite{GrafensteinCremer2000}. Unfortunately, in that approach, one cannot completely distinguish the dynamic and static contributions from complete active space of a variationally optimized wave function. As an alternative to that solution, one can introduce a dependence of exchange-correlation functional from active space directly through an appropriate quantity representing the amount of correlation accounted by multiconfigurational wave function \cite{GusarovRoos2004,GusarovLindh2004,ManniGagliardiOlsenTruhlar2014}. That way is theoretically more consistent but leads to complicated expressions for multiconfigurational DFT functionals which are difficult to parametrize. \\
This work presents a new approach and corresponding algorithm to avoid double counting problem by the decomposition of correlation energy into perturbation theory (PT) like series. Such attempts have been made in the past too \cite{Labzovsky1991}, but not with considerable success due to the complex structure of PT expressions. 
The consistent theoretical argumentation of the developed approach is similar to \cite{GusarovRoos2004,GusarovLindh2004} and below we present the details of computational methodology that utilizes  the Extended Koopmans' approximation. 

\section{Theory\label{Theory}}

For each $CASSCF$ active space $M$ we can construct a Green Function in the Extended Koopmans' approximation ($G^{EKT,M}$) by solving the generalized eigenvalue problem for the Koopmans' matrix (see \cite{GusarovCanJ2013,GusarovIJQC2009,GusarovIJQC2007,GusarovOptSp2007}). The $G^{EKT,M}$ is a much simpler object in comparison with the full one, because it has a simpler pole structure \cite{GusarovCanJ2013,GusarovIJQC2009,GusarovIJQC2007,GusarovOptSp2007}. However, it reproduces exact energy ($E^{CASSCF,M}$) and electron density ($\rho^{CASSCF,M}$) for selected active space M. Using $G^{EKT,M}$ and Dyson equation we can uniquely decompose $E^{CASSCF,M}$ into perturbation theory (PT) like series (see \cite{GusarovCanJ2013,GusarovIJQC2009,GusarovIJQC2007,GusarovOptSp2007}):
\begin{equation}
E^{CASSCF,M}=E^{CASSCF,M (0)} + E^{CASSCF,M (1)} + E^{CASSCF,M (2)} + ...
\label{eqn:PT}
\end{equation}
Of course, if $M_1 \subset M_2 \subset M_3 \subset ... \subset {FCI} $ 
then $E_1 > E_2 > E_3 > ... > E^{FCI} $ due to variational character of CASSCF approach but the hope is
that (and this is an approximation) starting from some reasonable $M_k$:
\begin{equation}
E^{CASSCF,M_k (0)} \approx E^{CASSCF,M_l (0)}, \qquad M_k \subset M_l .
\label{eqn:Approach}
\end{equation}
If this is true, we can define the new universal auxiliary functional:
\begin{equation}
\widetilde F^{CASSCF} [ \rho ] \equiv \lim_{M \to FCI} (E^{CASSCF,M } - E^{CASSCF,M (0)}) 
\label{eqn:F0}
\end{equation}
and then use it to construct the new CASDFT functional:

\begin{equation}
F^{CASSCF, M} [ \rho ] = \widetilde F^{CASSCF} [ \rho ] - (E^{CASSCF,M } - E^{CASSCF,M (0)}), 
\label{eqn:F}
\end{equation}
which depends on active space $M$. The expression in the parenthesis of (\ref{eqn:F}) represents the part of correlation energy already accounted by $CASSCF$ with active space $M$ and so should be subtracted from $\widetilde F^{CASSCF} $ to avoid double counting.

The above considerations can be summarized by the following algorithm:
\begin{enumerate}
\item Suppose we have the solution of CASSCF problem for reasonably small active space $M$ which
correctly represent an electronic structure. This resulted in the WF satisfying the BLB conditions,
$\Psi^{CASSCF, M}$  as well as density, $\rho^{CASSCF, M}$ and, CASSCF energy, $E^{CASSCF, M}$;
\item  Based on that solution, we can solve a generalized eigenvalue problem for Koopmans' matrix and then
construct one-particle Green's function in Extended Koopmans' approximation $G^{EKT,M} (\bf r)$ \cite{GusarovCanJ2013,GusarovIJQC2009};
\item  Next, we can decompose  $E^{CASSCF, M}$ into perturbation-like series \cite{GusarovCanJ2013,GusarovIJQC2009}
$E^{CASSCF,M (0)} + E^{CASSCF,M (1)} + E^{CASSCF,M (2)} + ...$ 
and construct a new EKT-CASDFT total energy: 
$E^{EKT-CASDFT, M} = E^{CASSCF, M} + F^{CASSCF, M} [ \rho ]$ which could be also represented as 
$E^{EKT-CASDFT, M} = E^{CASSCF, M (0)} + \widetilde  F^{CASSCF} [ \rho ]$  (but 
$F^{CASSCF, M} [ \rho ]$ should be also used to modify the CI-matrix);
\item Check for convergence if the termination conditions are satisfied and go to step 1 (if needed);
\end{enumerate}

The functional $F^{CASSCF, M}[\rho]$ in (\ref{eqn:F}) differs from traditional exchange-correlation $F_{xc}[\rho]$ in Kohn-Sham (KS) approach \cite{Kohn1965,Hohenberg1964} which is based on the idea of fictitious noninteracting particles moving in effective Kohn-Sham potential. But it will be very close to $F_{xc}[\rho]$ in the case if the one-determinant Hartree-Fock method is a good approximation to wave function. Moreover, the suggested new functional should have a better numerical behavior  because it is initially based on correct electronic structure (accounted by correct multiconfigurational CASSCF wave function for both parts of the difference in (\ref{eqn:F0})) and consequently does not depend on the correctness of the one determinant approximation. In other words, $E^{CASSCF,FCI (0)}$ should be a better starting point compare to one determinant approach because the multiconfigurational CASSCF wave function has the correct electronic structure. Moreover, by construction the new functional (\ref{eqn:F}) satisfies the criteria introduced in \cite{GusarovRoos2004,GusarovLindh2004}:
\begin{equation}
\lim_{M \to FCI} F^{CASSCF, M} [ \rho ] \to 0
\label{eqn:FCI}
\end{equation}
Based on the above, we further believe that EKT-CASDFT approach will be free of some sickness of traditional DFT. Moreover,  the majority of existing DFT functionals could be easily reparametrized to approximate $\widetilde F^{CASSCF}$  because it does not depend on any additional variables (e.g.$P_2$) which significantly simplifies its practical implementation.

\section{Conclusions\label{Conclusions}}

The developed approach extends the DFT theory to multiconfigurational wave function which allows to better account both static and dynamic correlation energy. Double counting of correlation energy is avoided by subtraction of the correlation energy accounted by CASSCF from universal functional $\widetilde  F^{CASSCF, M}[\rho]$. The details of construction and practical algorithm for implementation are presented. The practical aspects of implementation such as accuracy, convergence and computational cost are going to be studied in future works.

\begin{acknowledgments}
We would like to thank Dr. Andriy Kovalenko (NRC) and Prof. Per-{\AA}ke Malmquist (Lund University) for the insightful discussions.
\end{acknowledgments}

\bibliography{gusarov_ekt_casdft}

\end{document}